\documentstyle[aps,amsfonts,twocolumn,prl,floats,epsfig]{revtex}     

\begin{document}

\unitlength 1mm

\title{Long-time discrete particle effects versus kinetic theory
       in the self-consistent single-wave model}
\author{M-C.~Firpo$^{a,b}$\cite{byline}, F.~Doveil$^{a}$, Y.~Elskens$^{a}$,
        P.~Bertrand$^{c}$, M.~Poleni$^{c}$ and D.~Guyomarc'h$^{a}$}
\address{$^{a}$ Physique des interactions ioniques et mol{\'{e}}culaires,
                Unit{\'e} 6633 CNRS--Universit{\'e} de Provence, \\
                Equipe turbulence plasma,
                case 321, Centre de Saint-J{\'e}r{\^o}me,
                F-13397 Marseille cedex 20 \\
         $^{b}$ Dipartimento di energetica ``Sergio Stecco'',
                Universit\`{a} degli studi di Firenze, \\
                Via Santa Marta, 3, I-50139 Firenze \\
         $^{c}$ Laboratoire de physique des milieux ionis{\'{e}}s et
                applications,\\
                Unit{\'{e}} 7040 CNRS--Universit{\'{e}} H. Poincar{\'{e}}, Nancy I, \\
                BP 239, F-54506 Vand{\oe}uvre cedex
                }
\date{preprint TP00.09 - submitted for publication to {\bf Physical Review E}}
\maketitle

\begin{abstract}
The influence of the finite number $N$ of particles coupled to a
monochromatic wave in a collisionless plasma is investigated. For
growth as well as damping of the wave, discrete particle numerical
simulations show an $N$-dependent long time behavior resulting
from the dynamics of individual particles. This behavior differs
from the one due to the numerical errors incurred by Vlasov
approaches. Trapping oscillations are crucial to long time
dynamics, as the wave oscillations are controlled by the particle
distribution inhomogeneities and the pulsating separatrix
crossings drive the relaxation towards thermal equilibrium.
\newline %
PACS numbers: 05.20.Dd (Kinetic theory) \newline %
52.35.Fp (Plasma: electrostatic waves and oscillations) \newline %
52.65.-y (Plasma simulation) \newline %
52.25.Dg (Plasma kinetic equations) \newline
\end{abstract}

\pacs{PACS numbers: 05.20.Dd, 52.35.Fp, 52.65.-y, 52.25.Dg}

%



\section{Introduction}

It is tempting to expect that kinetic equations and their
numerical simulation provide a fair description of the time
evolution of systems with long-range or `global' interactions.
Yet, physical systems are obviously composed of a finite (albeit
large) number of degrees of freedom. For instance, a plasma is a
system of $N$ charged particles in interaction. A relevant issue,
both for fundamental and computational reasons, is to test whether
the kinetic description may hide truly physical, finite-$N$,
behaviors.

We address this issue for a typical example where long-range
interactions come into play, namely wave-particle interactions,
which are ubiquitous in the physics of hot plasmas
\cite{Antoni,Crawford,Castillo}.

Restricting to the electrostatic and collisionless case, it has
been shown that the interaction of resonant particles with
Langmuir long-wavelength modes can be described by a Hamiltonian
self-consistent model, instead of the traditional Vlasov-Poisson
kinetic approach. We consider the case where particles interact
with a single mode. In Section II, we present the single-wave
Hamiltonian model, whose universal features were recently
emphasized in different contexts \cite{Crawford,Castillo}. The
essential property of this model is its mean-field nature~:
resonant particles only interact with the wave, which is the only
effective collective degree of freedom representing longitudinal
oscillations of bulk particles. These bulk particles, whose
individual characters are absent from the physical mechanism at
work, namely wave-particle interaction, are then eliminated in our
approach. There are thus no direct particle-particle interactions
in our model. There is only a mean-field coupling, which enables
one to derive rather directly its kinetic limit \cite{Firpo}. We
use this single wave-particle model to investigate the
discrepancies between kinetic and finite-$N$ approaches.
Numerically, the finite-$N$ Hamiltonian microscopic dynamics is
computed through a symplectic code, whereas the numerical
implementation of the kinetic counterpart of this system involves
a semi-Lagrangian Vlasov solver.

In Section III, we focus on long time numerical simulations of
both kinetic and finite-$N$ systems for initial conditions
modeling bump-on-tail beam-plasma instability or damping. We
review the well-known results of linear theory and show how long
time behaviors are intrinsically related to the non-linear regime.
Our model offers a good alternative to address the recently
controversial issue \cite{Longtime,Brunetti} of the long-time
evolution encountered in the Vlasov-Poisson system. In Section IV,
we analyse quantitatively the finite-$N$ effects. In particular,
we show how the discrepancies between kinetic and finite-$N$ long
time behaviors can be related to the presently actively
investigated topics of the possible inadequacy of Gibbs
thermodynamics to predict time-asymptotic dynamical behaviors in
the limit $N \to \infty $ for {\it long-range} systems. More
precisely, we give hints to the non-commutation of limits $t \to
\infty $ and $N \to \infty $ for the mean-field model. We conclude
in Section V.

\section{Mean-field model for the Langmuir wave-particle interaction}

\subsection{Links between the traditional Vlasov-Poisson treatment of
Langmuir waves and the self-consistent wave-particle Hamiltonian model}

Without collisions, plasma dynamics is dominated by collective
processes. Langmuir waves and their familiar Landau damping and
growth \cite{Landau} are a good example of these processes, with
many applications, e.g. plasma heating in fusion devices and
laser-plasma interactions. For simplicity we focus on the
one-dimensional case, relevant to electrons confined by a strong
axial magnetic field, and assume that ions act as a neutralizing
fixed background. The traditional description of the interacting
particles and fields then rests on the (kinetic) coupled set of
Vlasov-Poisson equations \cite{ONeil65,warmbeam}. The current
debate on the long-time evolution of this system hints that
further insight in this fundamental process is still needed
\cite{Longtime}.

Our approach to (Langmuir) wave-particle interaction complements
this usual treatment in that, to capture the physical mechanism at
work, electrons are partitioned in two populations~: bulk and
tail. The idea behind this discrimination is simple~:
wave-particle interaction involves the resonant tail particles
whose velocity is close to the phase velocity of the wave under
consideration. These waves are just the collective macroscopic
degrees of freedom, capturing the longitudinal oscillations of
other non-resonant, bulk particles, so that these bulk particles
participate in the effective wave-particle dynamics only through
the waves.

Langmuir modes are thus collective oscillations of bulk particles,
represented by slowly varying complex amplitudes in an envelope
approximation. Their interaction with individual tail particles is
described by a self-consistent set of Hamiltonian equations
\cite{Antoni,Mynick,Tennyson}. These already provided an efficient
basis \cite{Drummond} for investigating the cold beam plasma
instability and exploring the nonlinear regime of the bump-on-tail
instability \cite{Doxas}. Analytically, they yield an intuitive
and rigorous derivation of spontaneous emission and Landau damping
of Langmuir waves \cite{Zekri}. Besides, as it eliminates the
rapid plasma oscillation scale $\omega_{{\rm p}}^{-1}$, this
self-consistent model offers a genuine tool to investigate
long-time dynamics.

\subsection{The single wave model}

We discuss the case of one wave interacting with the particles.
Though a broad spectrum of unstable waves is generally excited
when tail particles form a warm beam, the single-wave situation
can be realized experimentally \cite{Tsunoda} and allows leaving
aside the difficult problem of mode coupling mediated by resonant
particles \cite{Laval}. Moreover, recent studies
\cite{Crawford,Castillo} have stressed the genericness of the
single-wave model, which we discuss later on.

Consider an electrostatic potential perturbation %
$\Phi (z,\tau)=\phi_k(\tau )\exp i(kz-\omega_{k}\tau )+{\rm c.c.}$
(c.c. means complex conjugate) with complex envelope $\phi_k$, in
a one-dimensional plasma of length $L$ with periodic boundary
conditions (and neutralizing background). Wavenumber $k$
and frequency $\omega_{k}$ satisfy a dispersion relation %
$\epsilon(k,\omega_{k})=0$. The density of $N$ (quasi)-resonant
electrons is $\sigma (z,\tau )=(nL/N)\sum_{l=1}^{N}\delta
(z-z_{l}(\tau ))$, where $n$ is the electron number density and
$z_{l}$ is the position at time $\tau $ of electron labeled $l$
(with charge $e$ and mass $m$). Non-resonant electrons contribute
only through the definition of the dielectric function $\epsilon$,
so that $\phi_k$ and the $z_{l}$'s obey coupled equations
\cite{Antoni,Drummond}
\begin{eqnarray}
  \frac{d\phi_{k}}{d\tau }
  &=&
  {\frac {i n e}
         {\epsilon_0 k^ 2 N (\partial \epsilon /\partial \omega_k)}}
  \sum_{l=1}^{N} \exp [-ikz_l + i\omega_{k}\tau ]
  \label{phi} \\
  \frac{d^2 z_l}{d \tau^2}
  &=&
  {\frac {iek} m} \phi_{k} \exp [ikz_{l}-i\omega_{k}\tau ]
  + {\rm c.c.}
  \label{force}
\end{eqnarray}
with $\epsilon_{0}$ the vacuum dielectric constant. We renormalize
time and positions to $t = \alpha \tau $, $x_l = kz_{l} -
\omega_{k} \tau$ where $\alpha^3 = ne^{2}/[m\epsilon_{0}(\partial
\epsilon /\partial \omega_{k})]$. In particular, $\alpha
=(n/2n_{\rm p})^{1/3}\omega_{\rm p}$ for a cold plasma with
density $n_{\rm p}$, plasma frequency $\omega_{\rm p}$, and
dielectric function $\epsilon (k,\omega) = 1 - \omega_{\rm
p}^{2}/\omega^{2}$. With these rescaled variables and
$V=(ek^{2}\phi_{k})/(\alpha^{2}m)$, this system defines the
self-consistent dynamics (with $N+1$ degrees of freedom)
\begin{eqnarray}
  \dot V
  &=&
  iN^{-1}\sum_{l=1}^N \exp(-ix_l)
  \label{Vdot} \\
  \ddot x_l
  &=&
  iV\exp (ix_l)-iV^{*}\exp (-ix_l)
  \label{accel}
\end{eqnarray}
for the coupled evolution (in dimensionless form) of the electrons
and wave complex amplitude (a star means a complex conjugate and
${\dot { }} = d/dt$). This evolution derives from the Hamiltonian
\begin{equation}
  H(x_l,p_l,\zeta ,\zeta^{*})
  =
  \sum_{l=1}^{N} \left( \frac{p_l^{2}}{2}
                 -N^{-1/2}\zeta {\rm e}^{ix_l}
                 -N^{-1/2}\zeta^{*}{\rm e}^{-ix_l}
                 \right) ,
\label{ExpHam}
\end{equation}
where $\zeta = N^{1/2} V$. This system conserves energy ${\cal
H}=H$ and momentum ${\cal P}=\sum_{l}p_l+|\zeta |^{2}$. An
efficient fourth-order symplectic integration scheme is used to
study this Hamiltonian numerically \cite{Doxas}.

\subsection{Kinetic limit and study of finite-$N$ effects}

As we follow the motion of each particle, we can address the
influence of the finite number of particles on the long-time
behavior of the system. This question is eluded by the kinetic
Vlasov-Poisson description, and one might argue that finite $N$ is
analogous to numerical discretisation in solving kinetic
equations. Thus we investigate the kinetic limit, $N\to \infty $.
As there is no direct particle-particle interaction in our model
(\ref{ExpHam}), it is possible to express in a simple way the
$N\to \infty $ limit through a parallel treatment of the particles
and the wave. The mean-field coupling between collective (wave)
and individual (particles) degrees of freedom enables one to avoid
the derivation of a full BBGKY hierarchy.

In the kinetic limit $N\to \infty $, the discrete distribution
$\sigma $ is thus replaced with a density $f(x,p,t)$, and system
(\ref{Vdot})-(\ref{accel}) yields the Vlasov-wave system
\begin{equation}
  \frac{dV}{dt}=i\int \exp (-ix)f(x,p,t)dxdp  \label{Vkin}
\end{equation}
\begin{equation}
  \frac{\partial f}{\partial t}
  +p\frac{\partial f}{\partial x}
  +\left( iV\exp(ix)-iV^{*}\exp (-ix)\right) \frac{\partial f}{\partial p}
  =0.  \label{Vlakin}
\end{equation}
For initial data approaching a smooth function $f$ as $N \to
\infty$, the solutions of (\ref{Vdot})-(\ref{accel}) have been
proved to converge to those of the Vlasov-wave system
(\ref{Vkin})-(\ref{Vlakin}) over any finite time interval
\cite{Firpo}. This legitimizes our comparison between finite $N$
and kinetic behaviors.

The kinetic model (\ref{Vkin})-(\ref{Vlakin}) is integrated
numerically by a semi-Lagrangian solver, which covers the $(x,p)$
space with a rectangular mesh. The function $f$ (interpolated by
cubic splines) is transported along the characteristic lines of
the kinetic equation, i.e. along trajectories of the original
particles \cite{Bertrand}. Therefore, in addition to the truly
physical effects of the discrepancies between finite $N$ and
kinetic systems on long time simulations, we shall also compare in
this article computational finite grid effects of the kinetic
solver with the granular aspects of the $N$-particle system
\cite{DFlett}.

\subsection{Universal features of the single-wave model}

The single wave model was first formulated \cite{Drummond,Mynick}
as a simplified model to treat the instability due to a weak cold
electron beam in a plasma with fixed ions. For this singular case,
it was clear that retaining only a single Langmuir mode was a good
approximation even till some primary stage of non-linear
saturation. This derivation involved natural approximations, but
did not {\it a priori} preserve the Hamiltonian or Lagrangian
structure of the dynamics (though the latter is recovered in the
final equations), and a more direct derivation within the
Hamiltonian and Lagrangian formalism has been established
\cite{Antoni,Tennyson}.

Recently, different studies \cite{Crawford,Castillo} have extended
the regime of application of the single-wave model to a much
larger class of instabilities and have derived it in a generic way
in different contexts.

J.D. Crawford and A. Jayaraman \cite{Crawford} studied the
collisionless nonlinear evolution of a weakly unstable mode, in
the limit of a vanishing growth rate $\gamma \to 0^+$. They
derived in this limit, for a multispecies Vlasov plasma, the
asymptotic features of the electric field and distribution
functions. These reveal that the asymptotic electric field turns
out to be monochromatic (at the wavelength of the linear unstable
mode) and that the {\it nonresonant} particles respond to this
electric field in an essentially linear fashion whereas the {\it
resonant} particle distribution has a much more complicated
structure, determined by nonlinear processes, e.g. particle
trapping. That is, starting from a much wider class of
instabilities than the original single wave model proposed by
O'Neil, Winfrey and Malmberg \cite{Drummond}, Crawford and
Jayaraman derive asymptotic forms for the electric field and
distribution functions that precisely feature the assumptions for
the single wave model.

D. del-Castillo-Negrete \cite{Castillo} derived initially the
single wave picture using matched asymptotic methods to treat the
resonant and nonresonant particles. In the wider context of
self-consistent chaotic transport in fluid dynamics, he also
showed that the single-wave model provided a simplified starting
point to study the difficult problem of {\it active} transport (as
opposed to the transport of {\it passive} scalars which do not
affect the flow). Actually, the single-wave model captures the
essential features of the self-consistent transport of vorticity,
i.e. an advective field that modifies the flow while being
transported, through the constraint of a vorticity-velocity
coupling. Self-consistent active transport is a ubiquitous
phenomenon in geophysical flows or in fusion plasmas with the
problem of magnetic confinement.

Finally, we remark that the single wave model has close
connections, to be further clarified, with systems of coupled
nonlinear oscillators, such as those first studied by Kuramoto
\cite{Kuramoto}. The occurrence of a phase transition in the
regime of Landau damping \cite{FirpoTrans}, with order parameter
the mean-field intensity of the wave, is a manifestation of these
analogies.

Now we return to the original motivation of this work and review
wave-beam instability and damping.

\section{Wave-beam instability}

\subsection{Linear study}

Let us first study linear instabilities and remark that one
solution of (\ref {Vdot})-(\ref{accel}) corresponds to vanishing
field $V_{0}=0$, with particles evenly distributed on a finite set
of beams with given velocities. Small perturbations of this
solution have $\delta V = \delta V_{0} e^{\gamma t}$, with rate
$\gamma$ solving \cite{Zekri}
\begin{equation}
  \gamma
  =
  \gamma_{\rm r} + i \gamma_{\rm i}
  =
  iN^{-1} \sum_{l=1}^{N} (\gamma +i p_l)^{-2}.
  \label{disp}
\end{equation}
For a monokinetic beam with velocity $U$, (\ref{disp}) reduces to
$\gamma (\gamma +iU)^{2}=i$~; the most unstable solution occurs
for $U=0$, with $\gamma_{\rm r}=\sqrt{3}/2$ and $\gamma_{\rm
i}=1/2$. For a warm beam with smooth initial distribution $f(p)$
(normalized to $\int f dp = 1$), the continuous limit of
(\ref{disp}) yields
\begin{equation}
  \gamma = i\int (\gamma +ip)^{-2}f(p)dp.
  \label{diskin}
\end{equation}
For a sufficiently broad distribution ($|f^{\prime}(0)|\ll 1$), we
obtain $|\gamma_{\rm r}| \gamma_{\rm r}=\gamma_{\rm r} \pi
f^{\prime}(-\gamma_{\rm i})$, where $f^{\prime }=df/dp$, and, for
$|f^{\prime\prime }(0)|\ll \pi^{-1}$, one finds $\gamma_{\rm i}
\approx \pi \gamma_{\rm r} f^{\prime \prime }(0)$. Except for the
trivial solution given by $\gamma_{\rm r}=0$, one easily checks
that other solutions can only exist for a positive slope
$f^{\prime }(0)$. Then the perturbation is unstable as the time
behavior of $\delta V$ is controlled by the eigenvalue $\gamma $
with positive real part, i.e. with growth rate $\gamma_{\rm r}
\approx \gamma_{\rm L} = \pi f^{\prime }(0)>0$. For negative
slope, one recovers the linear Landau damping paradox
\cite{Landau}~: the observed decay rate $\gamma_{\rm L} = \pi
f^{\prime }(0)<0$ is not associated with genuine eigenvalues, but
with phase mixing of eigenmodes
\cite{Zekri,FirpoTrans,Firpo99,vKampenCase}. This is a direct
consequence of the Hamiltonian nature of the dynamics
\cite{Zekri}.

\subsection{Nonlinear regime}

This linear analysis generally fails to give the large time
behavior. This is obvious for the unstable case as non-linear
effects are no longer negligible when the wave intensity grows so
that the bounce frequency $\omega_{{\rm b}}(t)=\sqrt{2|V(t)|}$ of
trapped particles in the wave becomes of the order of the linear
growth rate $\gamma_{\rm r}$.



\begin{figure}[tbp]
  \centerline{
  \psfig{figure=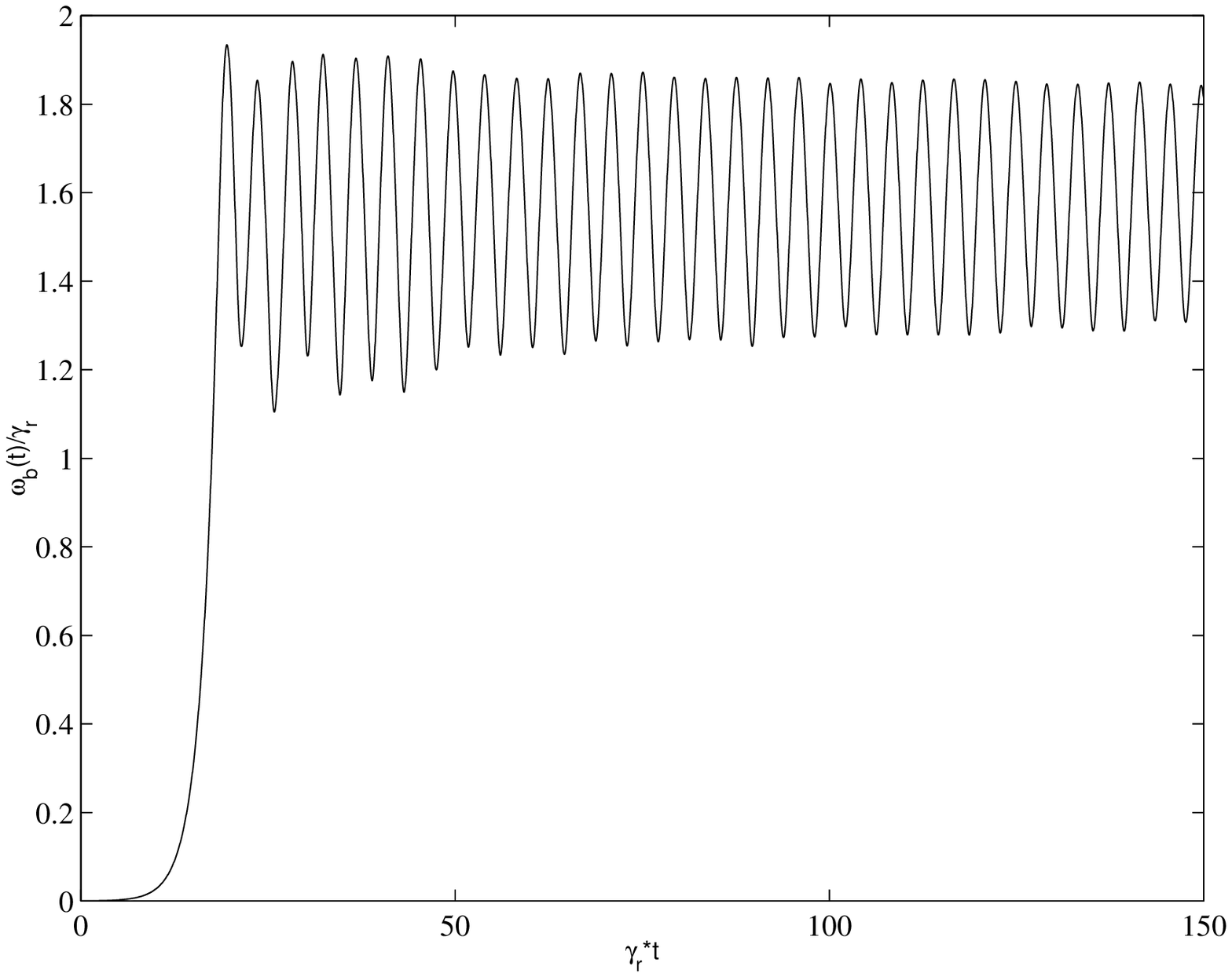,width=8cm,height=5cm}
  }
  \vskip5mm
\caption{ Time evolution of $\omega_{\rm b} (t) / \gamma_{\rm r}$
  for the cold beam instability. The initial velocity of the beam
  was chosen to maximize $\gamma_{\rm r}$ and the beam is
  initially spatially homogeneous.}
\label{fig001}
\end{figure}

\begin{figure}[tbp]
  \centerline{
  \psfig{figure=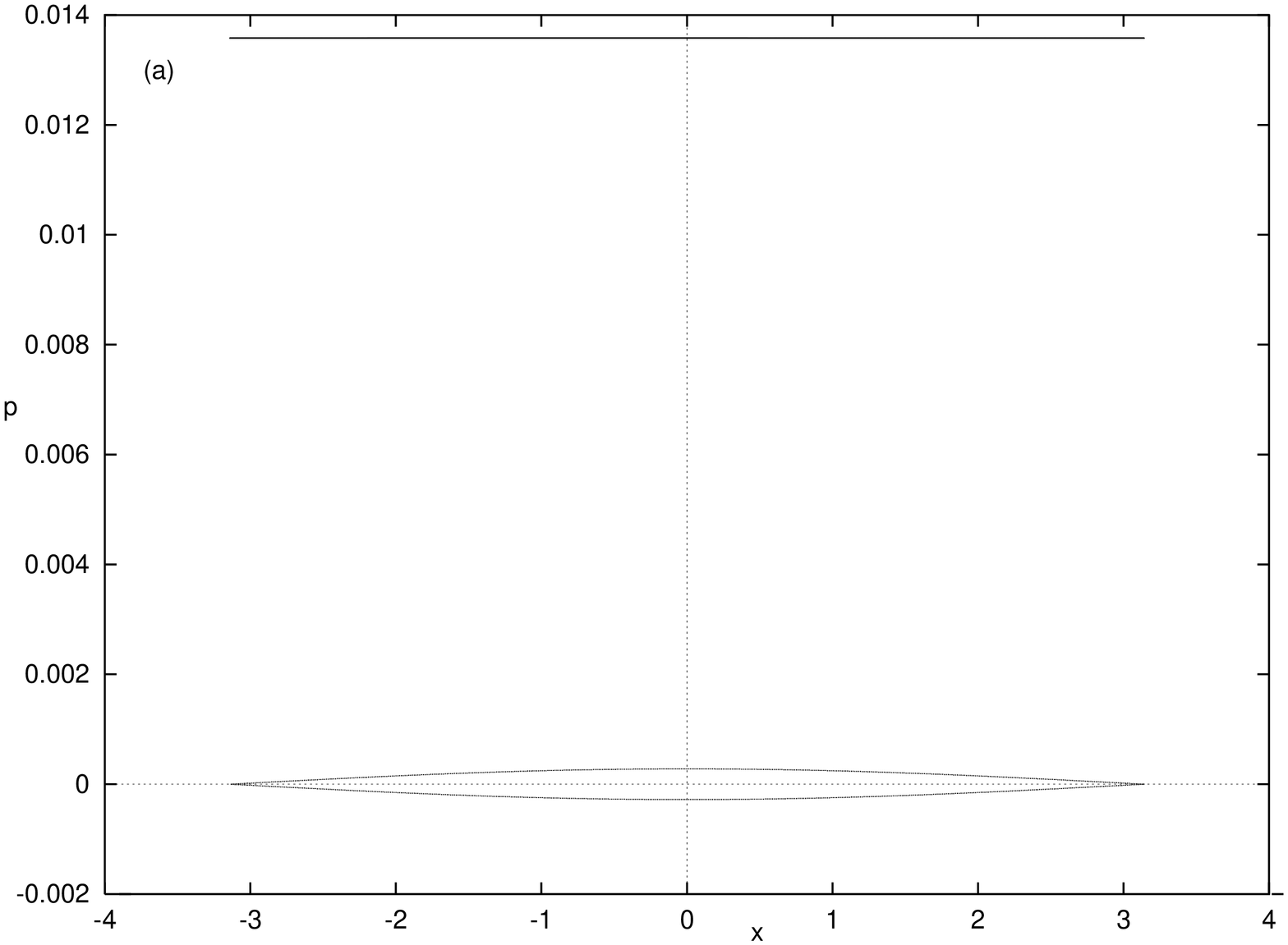,width=4cm,height=8cm,angle=-90}
  }
  \vskip1mm %
  \centerline{
  \psfig{figure=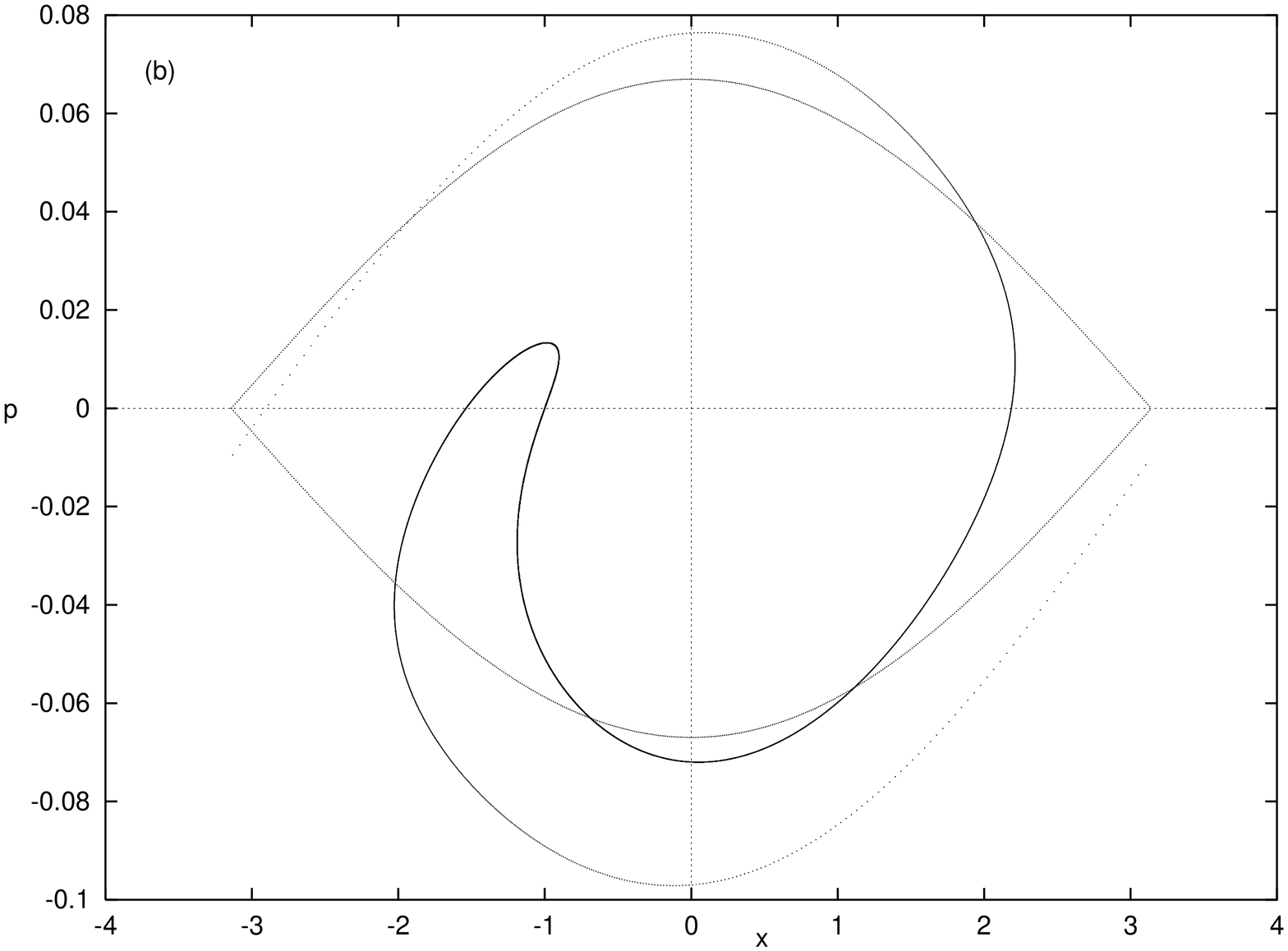,width=4cm,height=8cm,angle=-90}
  }
  \vskip1mm %
  \centerline{
  \psfig{figure=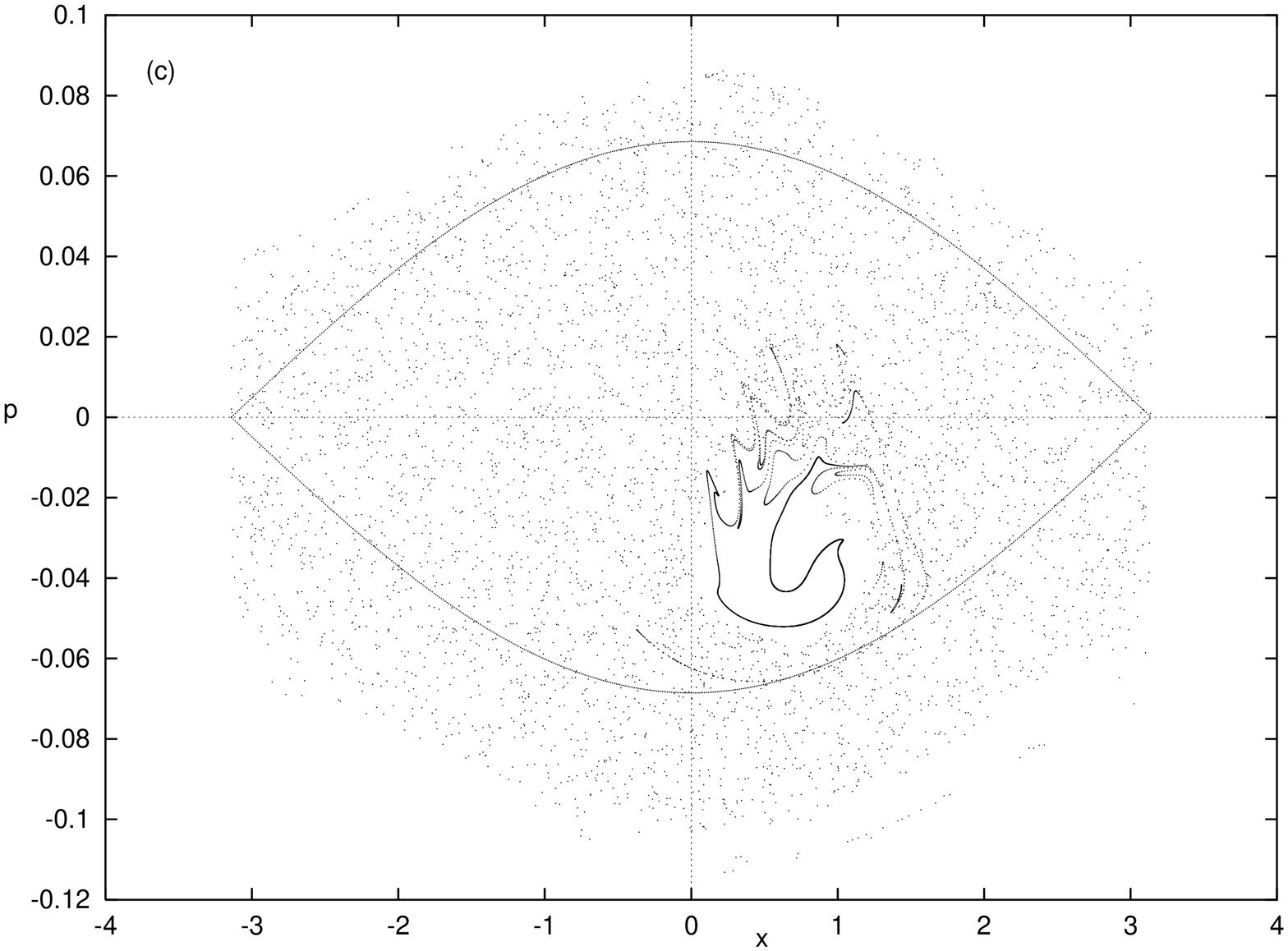,width=4cm,height=8cm,angle=-90}
  }
  \vskip5mm
\caption{ Snapshots of the $(x,p)$ space at
  (a)~$\gamma_{\rm r} t = 6.24$,
  (b)~$\gamma_{\rm r} t = 17.5$,
  (c)~$\gamma_{\rm r} t = 100$ for the cold beam simulation of
  Fig.~\ref{fig001}. Dots are the $N = 10000$ particles,
  which were initially distributed on a monokinetic beam,
  faster than the wave (with very small initial intensity).
  The instantaneous wave resonance `cat eye' is drawn to help
  in visualizing the instantaneous force on particles.}
\label{fig002}
\end{figure}

We used the monokinetic case as a testbed
\cite{Firpo99,Guyomarch}. As seen in Fig.~\ref{fig001}, finite-$N$
simulations show that the unstable solution grows as predicted,
until it saturates to a limit-cycle-like behavior where the
trapping frequency $\omega_{{\rm b}} (t)$ oscillates between $1.2
\gamma_{\rm r}$ and $2 \gamma_{\rm r}$. In this regime, some of
the initially monokinetic particles have been scattered rather
uniformly over the chaotic domain, in and around the pulsating
resonance, while others form a trapped bunch inside this resonance
(away from the separatrix) as observed in Fig.~\ref{fig002}
\cite{Guyomarch}. This dynamics is quite well described by
effective Hamiltonians with few degrees of freedom
\cite{Tennyson,Firpo99}. Note that it cannot be easily followed by
a numerical Vlasov solver, as the initial beam has a singular
velocity distribution function.

\begin{figure}[tbp]
  \centerline{
  \psfig{figure=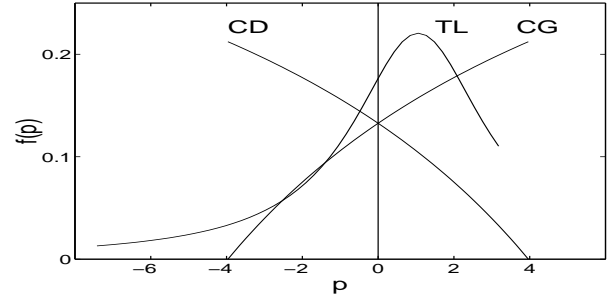,width=8cm,height=4cm}
  }
  \vskip5mm
\caption{ Initial warm beam velocity distributions.}
\label{fig003}
\end{figure}

In this article, we discuss the large time behavior of the warm
beam case, with $f^{\prime }(p_{0})\neq 0$ at the wave nominal
velocity $p_{0}=0$. Figure~\ref{fig003} displays three
distribution functions (in dimensionless form) with similar
velocity width (here $c$ normalizes $\int f dp = 1$ in each
case)~:

\noindent %
{\it (i)} a function (CD) giving the same decay rate for all phase
velocities, $f(p) = c - ac/(p_1 - p)$ if $-3.96 \leq p \leq 3.96$
and $f(p) = 0$ otherwise, with $a = 11.89$ and $p_1 = 15.85$,

\noindent %
{\it (ii)} a function (CG) giving a constant growth rate for all
phase velocities \cite{Doxas}, $f(p) = c - ac/(p_1 + p)$ if $-3.96
\leq p \leq 3.96$ and $f(p) = 0$ otherwise, with $a = 11.89$ and
$p_1 = 15.85$,

\noindent %
{\it (iii)} a truncated Lorentzian (TL) with positive slope $f(p)
= (c/\pi) / [ (p-p_1)^2 + a^2] $ if $ -7.42 \leq p \leq 3.18$ and
$f(p) = 0$ otherwise, with $a = 2.12$ and $p_1 = 1.06$.


For the growing case, nonlinearities result from the growth of the
wave intensity. For the damping case, the linear description
introduces time secularities which ultimately may cause the linear
theory to break down. The ultimate evolution is intrinsically
nonlinear, not only if the initial field amplitude is large, as in
O'Neil's seminal trapping picture for one wave \cite{ONeil65}, but
also if one considers the system evolution over time scales of the
order of the trapping time (which may be large if the initial wave
amplitude is small). The question of the long-time fate of plasma
wave amplitude is thus far from trivial \cite{Longtime}. Though
some simulations \cite{Feix} suggest that nonlinear plasma waves
eventually approach a Bernstein-Greene-Kruskal steady state
\cite{BGK} instead of a Landau vanishing field, the answer should
rather strongly depend on initial conditions \cite{Brunetti}. Our
$N$-particle, 1-wave system is the simplest model to test these
ideas.

\section{Finite $N$-effects and kinetic treatment in long-time dynamics}

First of all let us mention that for finite $N$, the particles are
initially distributed in $(x,p)$ so that their distribution
approaches $f$ smoothly in the large $N$ limit
\cite{Antoni,Doxas,Zekri}.

\subsection{The damping case}

\begin{figure}[tbp]
\centerline{
  \psfig{figure=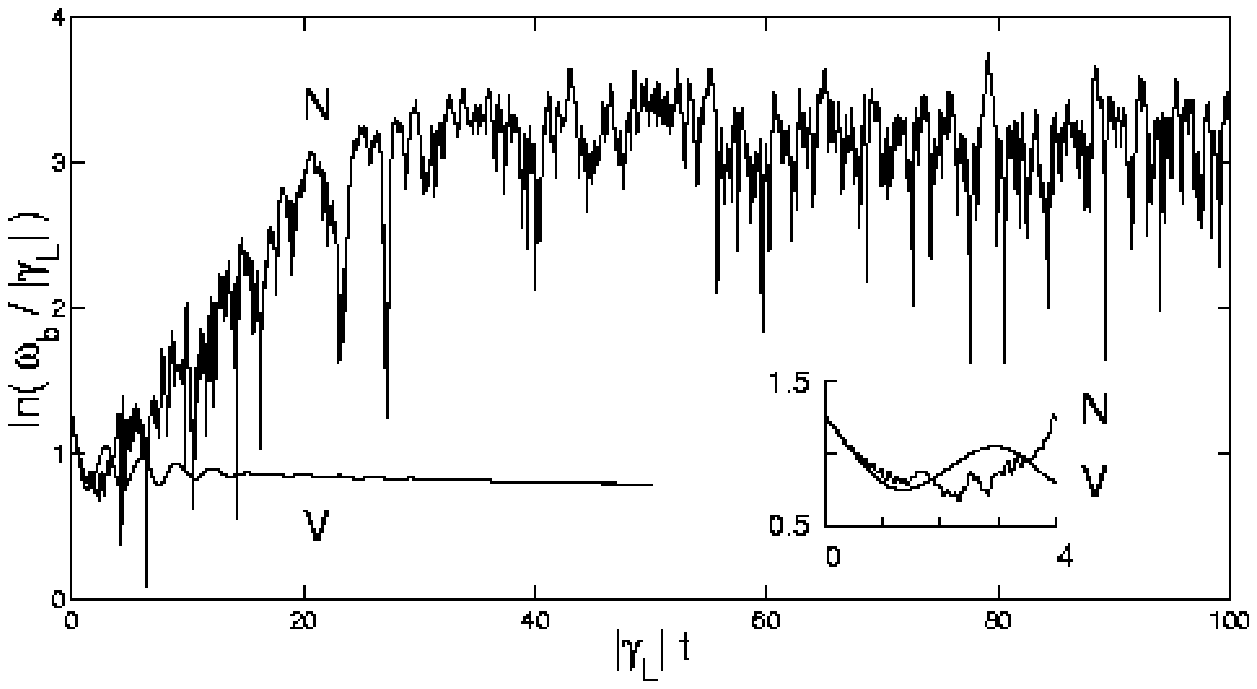,width=8cm,height=4cm}
  }
  \vskip5mm
\caption{ Time evolution of
  $\ln(\omega_{\rm b} (t) /|\gamma_{\rm L}|)$ for a CD velocity distribution
  and initial wave amplitude below thermal level~:
  (N)~$N$-particles system with $N=32000$,
  (V)~kinetic scheme with $32 \times 512$ $(x,p)$ grid.
  Inset~: short-time evolution.}
\label{fig004}
\end{figure}

A thermodynamical analysis \cite{FirpoTrans} predicts that, for a
warm beam (i.e. if the velocity distribution $f$ has a large
width, as in Fig.~\ref{fig003}) and small enough initial wave
amplitude, $\omega_{{\rm b}}$ asymptotically scales as $N^{-1/2}$
in the limit $N\to \infty$. Figure~\ref{fig004} shows the
evolution of a small amplitude wave launched in the beam. The
$N$-particle system (curve N) and the kinetic system (curve V)
initially damp exponentially as predicted by perturbation theory
\cite{Zekri}, for a time of the order of $|\gamma_{\rm L}|^{-1}$.
After that phase-mixing time, trapping induces nonlinear evolution
and both systems evolve differently. For the $N$ -particle system,
the wave grows to a thermal level that scales as $N^{-1/2}$,
corresponding to a balance between damping and spontaneous
emission \cite{Zekri,FirpoTrans}. For the kinetic system, initial
Landau damping is followed by slowly damped trapping oscillations
around a mean value~; this mean value also decays to zero, at a
rate which decreases for refined mesh size. Figure~\ref{fig004}
thus reveals that finite $N$ and Vlasov behaviors can considerably
diverge as spontaneous emission is taken into account.



\begin{figure}[tbp]
  \centerline{
  \psfig{figure=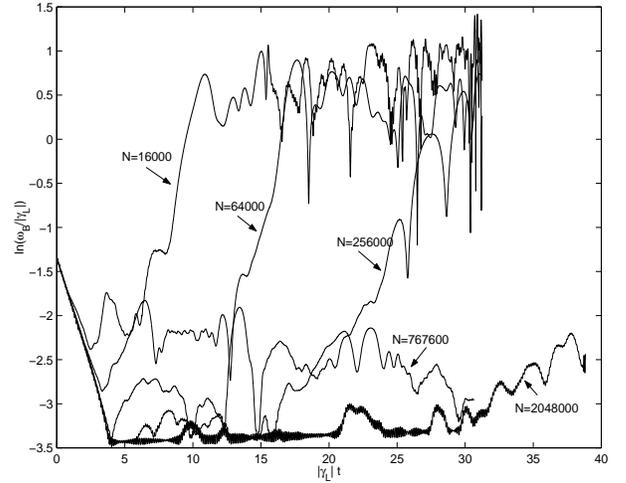,width=8cm,height=6.5cm}
  }
\caption{ Time evolution of
  $\ln (\omega_{\rm b}(t)/|\gamma_{\rm L}|)$ for an initial
  CD velocity distribution and different values of $N$.
  When necessary for readability, curves were
  smoothed through a sliding window average.}
\label{fig005}
\end{figure}

Figure~\ref{fig005} represents the time evolution of the wave
amplitude for different values of $N$. It clearly shows how
finite-$N$ wave evolutions depart progressively from the $N\to
\infty$ curve (the later for larger $N$). One should also notice,
from Figs~\ref{fig004} and \ref{fig005} and for sufficiently large
$N$, the onset of nonlinear effects at large time. In spite of the
smallness of the initial values of the wave amplitude, nonlinear
effects (through trapping) eventually come into play and stop
Landau exponential decay, marking the beginning of a different
dynamical regime for which the decay is far slower.

\subsection{The single wave warm beam instability}

\begin{figure}[tbp]
  \centerline{
  \psfig{figure=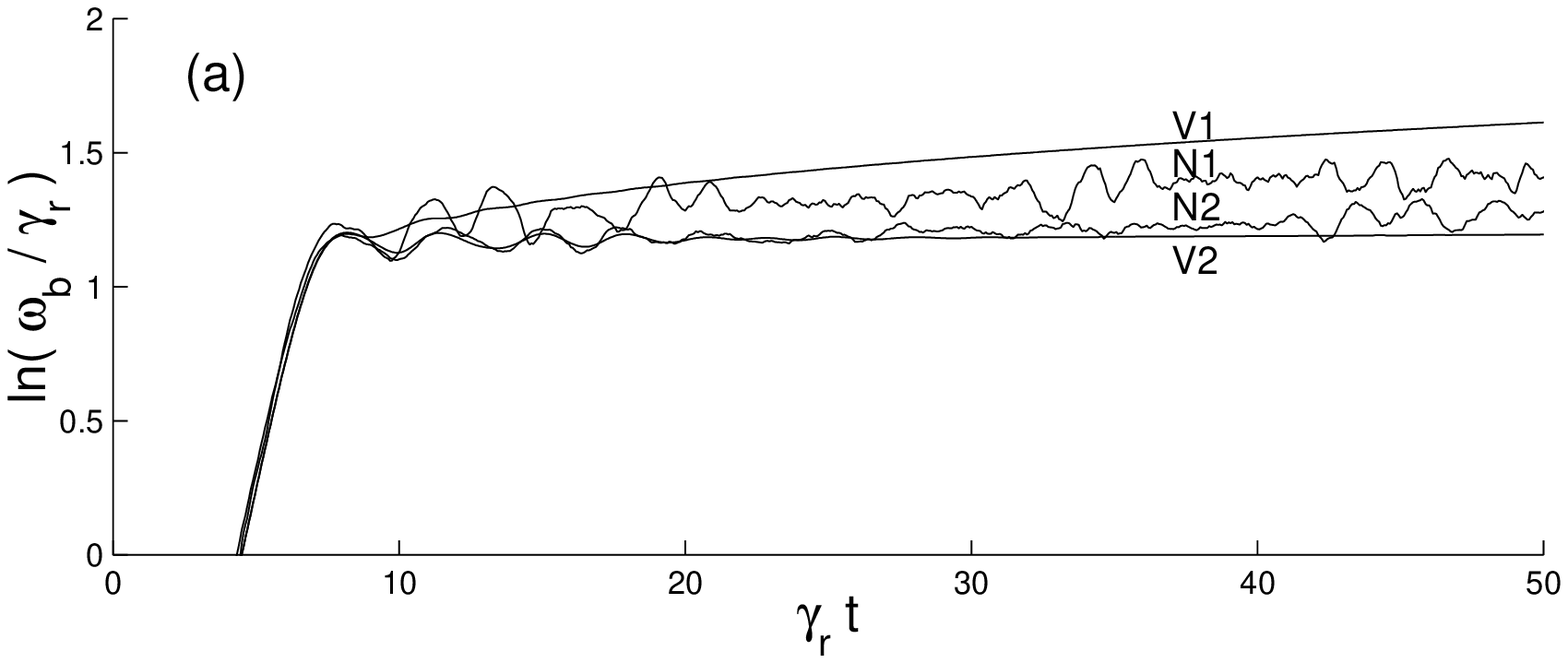,width=8cm,height=3.5cm}
  }
  \vskip1mm %
  \centerline{
  \psfig{figure=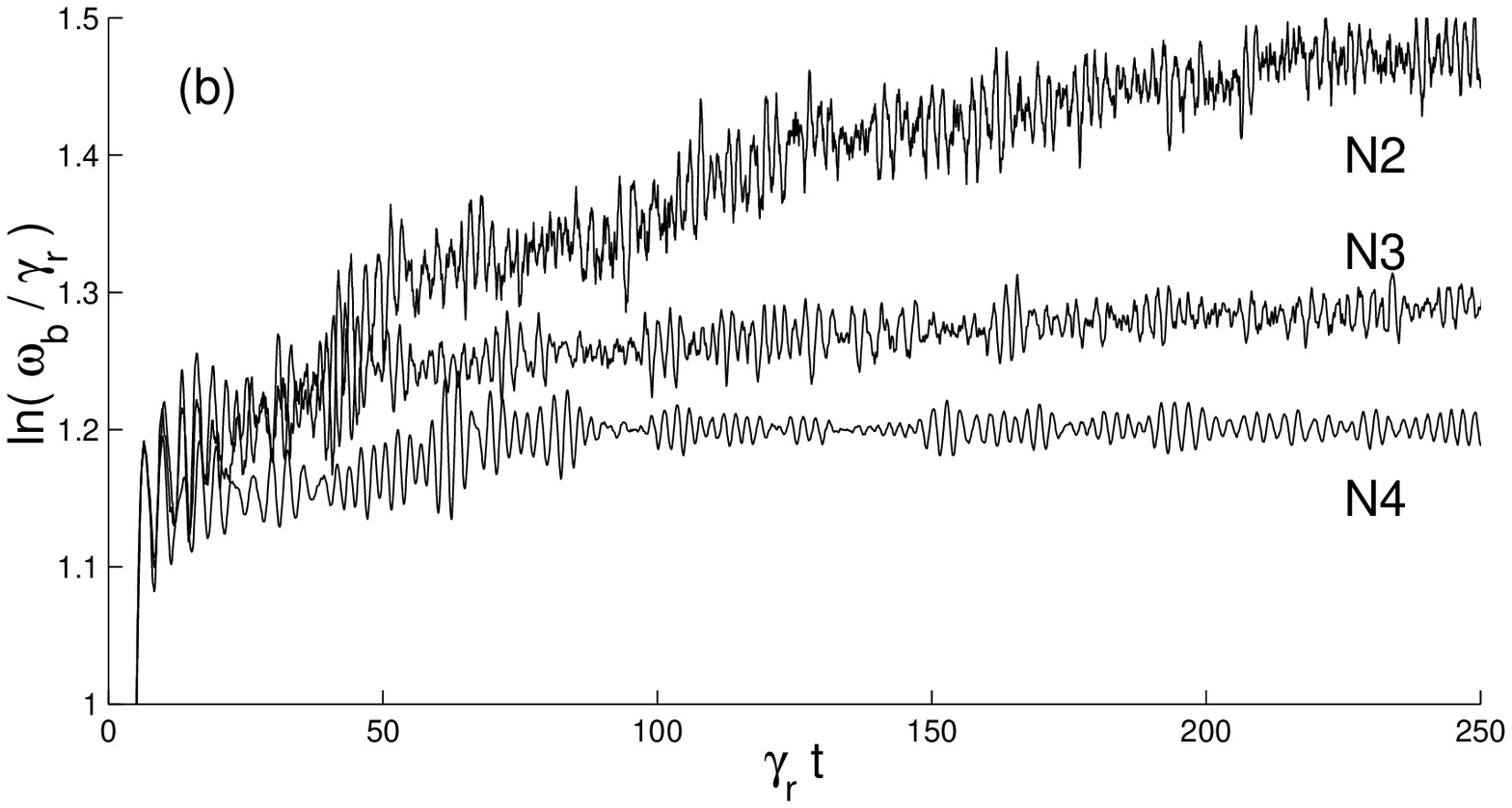,width=8cm,height=4cm}
  }
  \vskip5mm
\caption{ Time evolution of
  $\ln(\omega_{\rm b} (t) /\gamma_{\rm r})$.
  (a)~CG initial distribution~: kinetic scheme with
  (V1)~$32 \times 128$, (V2)~$256 \times 1024$ $(x,p)$ grid ;
  $N$-particles system with (N1)~$N=128000$, (N2)~$N=512000$~;
  (b)~Comparison of CG~(N2) with TL initial distribution for
  (N3)~$N=64000$, (N4)~$N=2048000$.}
\label{fig006}
\end{figure}

Now consider a warm beam with a velocity distribution with a
positive slope at $p_0=0$. Line N1 (resp. N2) of Fig.~\ref{fig006}
displays $\ln(\omega_{\rm b} (t) /\gamma_{\rm r})$ versus time in
a numerical integration of (\ref {Vdot})-(\ref{accel}) for a CG
distribution with $N=128000$ (resp. 512000) and $\gamma_{\rm r} =
0.08$. Line V1 (resp. V2) shows the evolution of $\ln(\omega_{\rm
b} (t) /\gamma_{\rm r})$ versus $\gamma_{\rm r} t$ in numerical
integration of the kinetic system for the same initial conditions
with a $32 \times 128$ (resp. $256 \times 1024$) grid in $(x,p)$
space. All four curves exhibit the same initial exponential growth
of linear theory with less than 1\% error on the growth rate.
Saturation occurs for $\omega_{\rm b} / \gamma_{\rm r} \approx
3.1$ \cite{warmbeam}. Lines N1 and V1 do not superpose beyond the
first trapping oscillation after saturation. Note that, in our
system, oscillating saturation cannot be related to excitation of
sideband or satellite Langmuir waves as our single-wave
Hamiltonian does not allow for any spectrum of waves.

Beyond the first trapping oscillation, kinetic simulations exhibit
a second growth at a rate controlled by mesh size. Line V2
suggests that a kinetic approach would predict a level close to
the trapping saturation level on a time scale awarded by
reasonable integration time. This level is fairly below the
thermodynamic level $V_{\rm th}$ predicted by a Gibbsian approach
\cite{FirpoTrans}. Such pathological relaxation properties in the
$N \to \infty $ limit seem common to mean-field long-range models
\cite{Latora}. Both kinetic simulations also exhibit a strong
damping of trapping oscillations, which disappear after a few
oscillations, whereas finite-$N$ simulations show persistent
trapping oscillations.


One could expect that finite-$N$ effects would mainly damp these
oscillations, so that the wave amplitude reaches a plateau.
However their amplitude does not depend on the number of particles
$N$, which shows that they are not an artifact due to `poor
accuracy' of finite-$N$ simulations. Moreover the wave amplitude
slowly grows further, whereas the velocity distribution function
flattens over wider intervals of velocity \cite{DFlett,Guyomarch}.

This second growth after the first trapping saturation depends on
the shape of the initial distribution function. In
Fig.~\ref{fig006}(b), curve N2 is the same as in
Fig.~\ref{fig006}(a) but computed over a longer duration, and
curve N3 corresponds to $N = 64000$ with the TL distribution of
Fig.~\ref{fig003}. Although curve N3 corresponds to 8 times fewer
particles than curve N2, the final level reached at the end of the
simulation is lower. In the second growth regime, particles are
transported further in velocity, so that the plateau in $f(p)$
broadens with time. As will be clearly shown in Sec.~D, this
spreading is due to separatrix crossing, i.e. successive trapping
and detrapping by the wave \cite{Guyomarch}. As the resonance
width of the wave separatrix grows, the wave can trap particles
with initial velocity further away from its phase velocity. Noting
that the TL distribution decays while the CG distribution still
grows for $v > 0.05$, we see that, with TL, fewer particles can
give momentum to the wave when being trapped (as ${\cal P}$ is
conserved)~; hence the second growth is slower for the TL
distribution.

We followed the evolution of the wave amplitude of curve N3 of
Fig.~\ref{fig006}(b) up to $\gamma_{\rm r}t=1750$. Starting from
the first trapping saturation level, equal to 40\% of the
thermodynamic level $V_{\rm th}$, we observe persistent amplitude
fluctuations with a growth rate that slowly decreases as we reach
$0.78V_{\rm th}$ at the end of the computation.

Line N4 of Fig.~\ref{fig006}(b) corresponds to the TL distribution
with 2048000 particles and shows persistent oscillations with
approximately the same amplitude as for $N=64000$.

\subsection{Trapping oscillations}

Let us show that the occurrence of trapping oscillations with
non-vanishing amplitude follows from the existence of spatial
inhomogeneities. For this purpose, introduce the complex field
\begin{equation}
  M^{(N) }
  =
  \frac{1}{N} \sum_{l=1}^{N} \exp (ix_l)
  \equiv
  \left| M^{(N) }\right| \exp (i\alpha ^{(N) }).
  \label{defMN}
\end{equation}
In the kinetic Vlasov limit, it is the first Fourier component of
the spatial density
\begin{equation}
  M^{(\infty) }
  =
  \int \exp(ix) f(p,x,t) dpdx.
  \label{defMkin}
\end{equation}
A spatially homogeneous phase space with independent particles
corresponds obviously to a $\left| M^{(N)}\right| $ scaling as
$N^{-1/2}$ i.e. to a vanishing $ \left| M^{( \infty ) }\right| $.
From (\ref{Vdot}), and dropping the superscript $(N)$, it follows
that
\begin{equation}
  \dot{V}=iM^*.
  \label{relVdotM}
\end{equation}
Putting
\begin{equation}
  2 V
  =
  2 \left| V \right| \exp (-i \theta)
  =
  \omega_{\rm b}^{2} \exp (-i \theta ),
  \label{Vcomplex}
\end{equation}
one obtains from (\ref{relVdotM})
\begin{equation}
  \frac{d |V| }{dt}
  =
  |M| \sin (\alpha - \theta ).
\label{rel1}
\end{equation}
Moreover, if the wave amplitude displays a sinusoidal temporal
evolution (e.g. due to trapping oscillations, then $\omega_1 =
\omega_{{\rm b}0}$) such that
\begin{equation}
  |V| (t)
  =
  {\frac 1 2} \omega_{{\rm b}0}^{2}
  +
  \Delta V \sin ( \omega_1 t )
  \label{modeltrap}
\end{equation}
with $\Delta V>0$ the amplitude of oscillations and $\omega_{{\rm
b} 0}$ the (quadratic) average bounce pulsation, one obtains
\begin{equation}
  \frac{d |V| }{dt}
  =
  \omega_1 \Delta V \cos ( \omega_1 t)
  \label{rel2}
\end{equation}
so that taking the time average, denoted by $\langle \cdot
\rangle_{t}$, of the square of both (\ref{rel1}) and (\ref{rel2})
one gets
\begin{equation}
  \Delta V
  =
  \omega_1^{-1} \left\langle | M |^{2} \right\rangle_{t}^{1/2}
\label{relatrap1}
\end{equation}
provided that $\alpha - \theta$ has a uniform distribution (e.g.
if $\alpha - \theta \approx \omega_2 t$ for some $\omega_2$). This
simplified model, supposing only a harmonic oscillation for $|V|$,
shows that the amplitude of the wave oscillations depends directly
on the occurrence of $(x,p)$-space inhomogeneities. Actually, for
a homogeneous phase space, (\ref{relatrap1}) implies that $\Delta
V$ scales at most as $N^{-1/2}$ and vanishes in the kinetic limit.

However, consider now the case where $\alpha - \theta$ is the
barycentric position, in the reference frame of the wave, of an
inhomogeneity (clump) composed of a finite fraction $| M |$ of the
particles. If this clump is sufficiently close to the bottom of
the potential well, then $\alpha(t) - \theta(t)$ is small and
(\ref{relatrap1}) can be reduced to
\begin{equation}
   \Delta V
   =
   \omega_1^{-1} | M |
   \langle 2 (\alpha - \theta)^{2} \rangle_{t}^{1/2},
\label{relatrap2}
\end{equation}
where $\left\langle (\alpha - \theta)^{2} \right\rangle_{t}^{1/2}$
is the mean distance of the clump from the center of the resonance
cat's eye. Note that $\Delta V$ may be small even if $M$ is large,
provided that the clump stays close to the bottom of the well
\cite{NonexistenceNote}.

A striking example is given by the cold beam-wave instability
\cite{Tennyson}, for which a macroscopic fraction of the particles
belong to a so-called {\it macroparticle} that oscillates near the
bottom (elliptic point) of the potential well and drives the wave
amplitude oscillations as shown in Fig.~\ref{fig002}. Another
illustration of (\ref{relatrap2}) is given by Figs~2 and 4 of
ref.~\cite{Brunetti}. In Vlasov simulations, as the spatial
resolution $N_{x}$ is increased, one observes a more refined phase
space that reveals some heterogeneities for the larger value of
$N_{x}$. This simultaneously goes with pronounced trapping
oscillations at large $N_{x}$ while the rough resolution, which
has smeared out the thin filamentation in the vortices, is
associated to a flat, constant amplitude in time.

\subsection{Finite-$N$ effects, trapping oscillations and relaxation towards
thermal equilibrium through chaos} \label{FiniteN}

We now discuss the actual process by which the wave-particle
system relaxes. For this purpose, one can get a flavor of the
stochasticity (strong or weak depending on resonance overlapping
or not) that a test particle would encounter in $(x,p)$ space at
different stages of the evolution.

In the equation of motion (\ref{accel}) of any particle in the
system (\ref{ExpHam})
\begin{equation}
  \ddot x
  =
  - \omega_{\rm b}^2 (t) \sin ( x - \theta(t) )
  \label{mouvpartest}
\end{equation}
the time dependence of the wave modulates the force on the test
particle. If $V$ is approximately periodic over a time interval of
length $T$, and if its period is large compared with the trapped
particle bouncing period, the pulsations of the separatrix
generate strong chaos in the domain of $(x,p)$ space it sweeps
\cite{slowchaos}. In the present case, the period of $V$ is
basically the trapped particles' bouncing period, which makes the
`slow chaos' approximation rather crude.

The chaos generated by a pulsating separatrix can also be
characterized using the Fourier decomposition of the wave $V(t) =
\sum_{\omega_n=-\infty}^{\infty} |v_n| \exp(i \omega_n t +
\chi_n)$, with $\omega_n = 2 \pi n /T$. Then the test particle
experiences a force deriving from an {\it effective many-wave
field} (though the Hamiltonian (\ref{ExpHam}) involves a single
wave),
\begin{equation}
  \ddot x
  =
  -2 \sum_n |v_n| \sin(x + \omega_n t - \chi_n)
  \label{EffAcc}
\end{equation}
and the overlaps between resonances in this force field cause the
particle to move chaotically \cite{overlap}.

\begin{figure}[tbp]
  \centerline{
  \psfig{figure=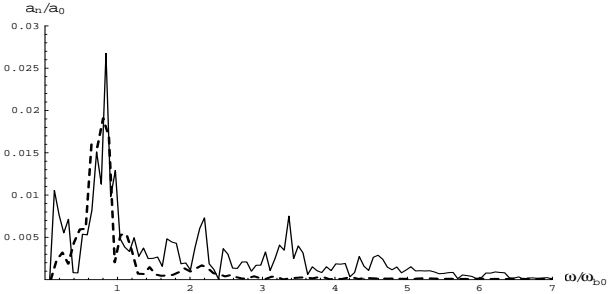,width=8cm,height=4cm}
  }
  \vskip5mm
\caption{ Coefficients $a_n / a_0$ ($n \geq 1$) of the Fourier
  decomposition of amplitude $\omega_{\rm b}^2 (t)$ during the
  first time window $38 \leq \gamma_{\rm L} t \leq 65$
  for $N=48000$ (solid curve) and $N=768000$ (dashed curve)
  particles, as a function of normalized frequency.}
\label{fig007}
\end{figure}

\begin{figure}[tbp]
  \centerline{
    \psfig{figure=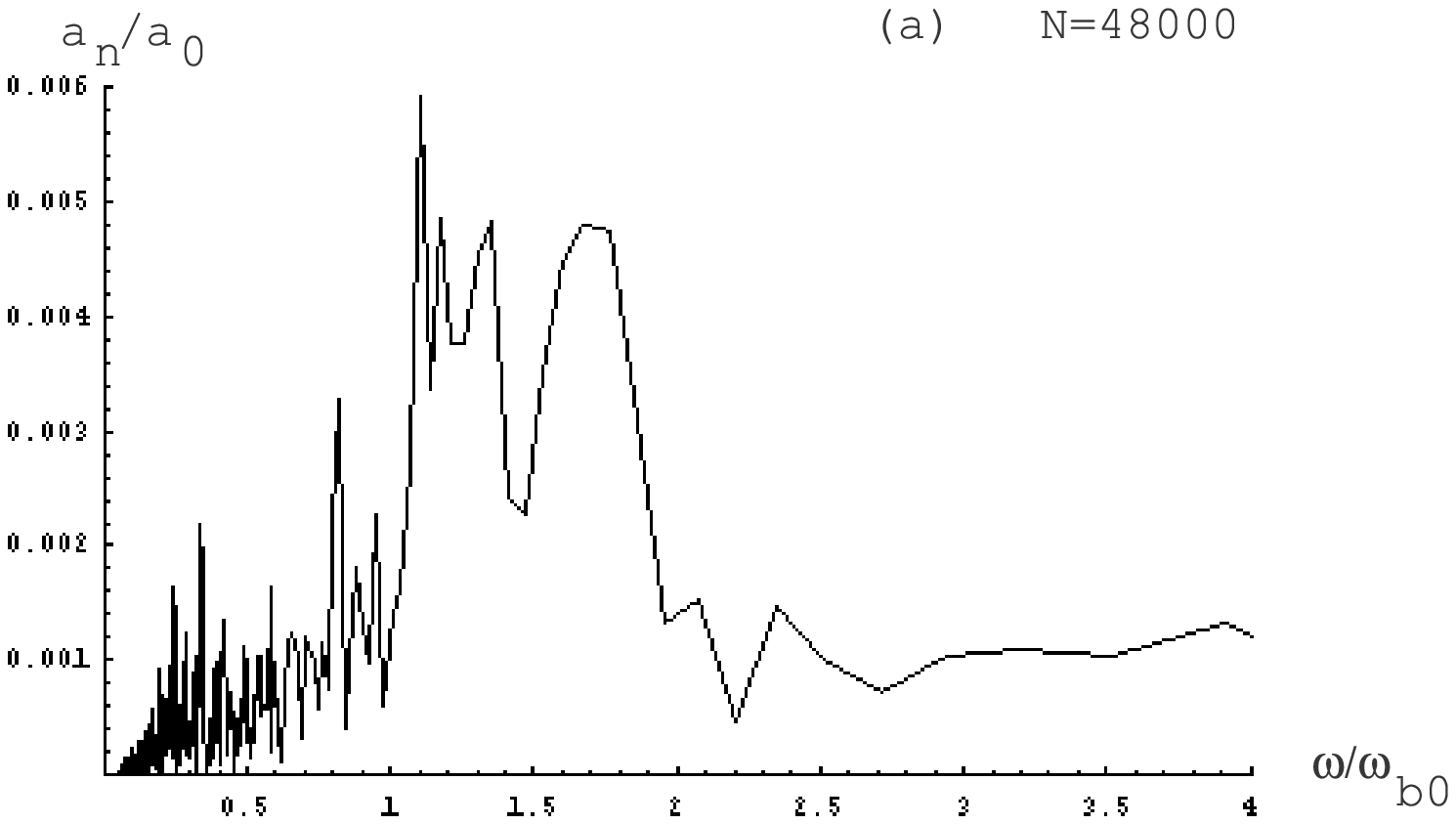,width=8cm,height=4cm}
    }
  \centerline{
    \psfig{figure=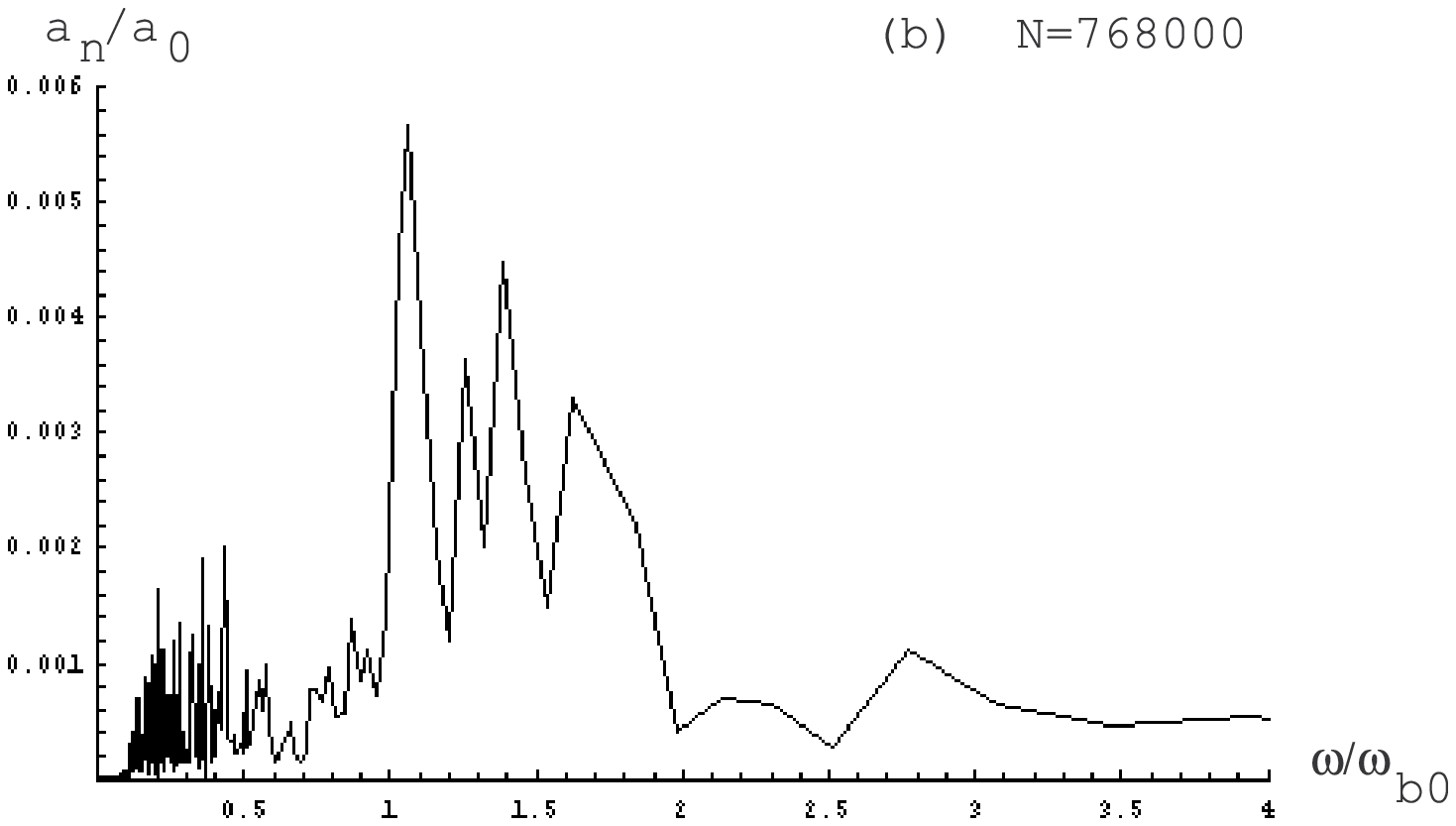,width=8cm,height=4cm}
    }
  \vskip5mm
\caption{ Coefficients $a_n / a_0$ ($n \geq 1$) of the Fourier
  decomposition of amplitude $\omega_{\rm b}^2 (t)$ during the
  second time window $250 \leq \gamma_{\rm L} t \leq 302$
  as a function of normalized frequency, for
  (a)~$N=48000$ and for
  (b)~$N=768000$ particles.}
  \label{fig008}
\end{figure}

We computed the Fourier decomposition of $\omega_{\rm b}(t)$ for
the warm beam with initial distribution TL over two different time
intervals, one (window $T_1$) just after the nonlinear saturation
with $38 \leq \gamma_{\rm L} t \leq 65$ and the other one (window
$T_2$) far later with $ 250 \leq \gamma_{\rm L} t \leq 302$, for
$N=48000$ and $N=768000$ particles. The low frequency bias induced
by the slow growth of the amplitude evolution has been removed by
subtracting the quadratic best fit of the wave evolution. The
Fourier decomposition (with $2\pi /T \ll \omega_{{\rm b}0}$ and
$T=T_{1}$ or $T_{2}$) reads
\begin{equation}
  \omega_{\rm b}^{2} (t)
  =
  a_0 \left( 1
           + \sum_{n=1}^\infty {\frac {a_n} {a_0}}
                               \cos( \omega_n t + \varphi_n)
           \right)
  \label{omegaFourier}
\end{equation}
where
\begin{equation}
  a_0
  =
  \omega_{{\rm b}0}^2
  =
  {\frac 1 T} \int_{t_0}^{t_0 + T} \omega_{\rm b}^2 (t) dt
\end{equation}
with $\gamma_{\rm L} t_0 = 38$ or $250$. Figs~\ref{fig007} and
\ref{fig008} show coefficients $a_n/a_0$ as a function of the
frequency normalized by $\omega_{{\rm b}0}$, the fundamental $n=0$
being removed. Such a normalized form enables a direct comparison
between the figures (irrespective of $N$ and $T$). Below we use
coefficients $a_{{\rm b}n} = a_{n'}$, where $n'$ is the index such
that $\omega_{n'} = n' 2 \pi / T = n \omega_{{\rm b}0}$.

During the first stage $T_1$ (Fig.~\ref{fig007}), the behavior
appears almost entirely driven for large $N$ by a narrow spectrum
of frequencies of the order of the average trapping frequency,
although, for $N=48000$, other Fourier components are excited.

Two types of chaos must be distinguished. The simplest one is
related to the classical overlap between the resonances of two
waves \cite{overlap} propagating at velocities $m \omega_{{\rm
b}0}$ and $n \omega_{{\rm b}0}$. The corresponding chaos is
`fast', namely the frequencies of the relevant resonances are
larger than $\omega_{{\rm b}0}$. The stochasticity parameter
\cite{ForceNote} estimated as
\begin{equation}
  s_{m,n}
  =
  \frac {\sqrt{2 a_{{\rm b}n}} + \sqrt{2 a_{{\rm b}m}}}
        {|m - n| \omega_{{\rm b}0}}
  =
  \frac
     { \sqrt{2 a_{{\rm b}n} / a_0} + \sqrt{2 a_{{\rm b}m} / a_0}
     }
     {|m - n|}
\label{stoch01}
\end{equation}
is small for all $m \neq 0$, $n \neq 0$. This means that the
effective many-wave field does not generate strong chaos in
velocity ranges away from the wave nominal velocity.

Similarly, one may search for chaos induced by the overlap between
the component in (\ref{EffAcc}) with phase velocity $\omega_{{\rm
b} n} > 2 \omega_{{\rm b}0}$, and the central resonance, with
phase velocity 0. Their resonance overlap \cite{ForceNote}
parameter $s_{0,n} = (2 \sqrt{a_0} + \sqrt{2 a_{{\rm b}n}})/|n
\omega_{{\rm b}0}| \approx 2/n$ is not large enough in our system
to induce large scale chaos, transporting particles from the
neighborhood of the wave nominal velocity (with its natural width
$2 \omega_{{\rm b}0}$) to the neighborhood of waves with
significantly different phase velocities.

The second type of chaos is related to the resonant forcing by
Fourier components with phase velocity $\omega_{{\rm b} n}
\lesssim 2 \omega_{{\rm b}0}$, on the periodic motion induced by
the main field component. This process applies in the vicinity of
the separatrix of the main resonance and must be discussed using
action-angle variables. As recalled in the appendix, the
corresponding pulsations are smaller than $2\omega_{{\rm b}0}$ for
untrapped particles, and smaller than $\omega_{{\rm b}0}$ for
trapped particles (and for untrapped ones very close to the
separatrix). Particles experiencing this chaos easily cross the
pulsating separatrix, i.e. change between trapped and untrapped
motion.

The Fourier spectra in Fig.~\ref{fig007} show that the second
process is active, since only $\omega \lesssim \omega_{{\rm b}0}$
lead to significant amplitudes $a_{{\rm b}n}$. This supports the
analysis of chaos in our system as `slow chaos' due to the
pulsating resonance \cite{slowchaos}. However, corresponding
amplitudes $a_{{\rm b}n}$ are much smaller than $a_0$. Therefore,
the pulsating resonance does not sweep the vicinity of the bottom
of the wave potential well, which allows the particles close to
the elliptic point to move quite regularly, forming a
macroparticle like in Fig.~\ref{fig002}. This bottom of the well
may be separated from the surrounding chaotic domain (swept by the
separatrix) by KAM surfaces in $(x,p,t)$ space \cite{slowchaos}.

The more chaotic behaviour is expected in the case where the
additional peaks, close to $\omega_{{\rm b}0}$, $2 \omega_{{\rm
b}0}$ and $3 \omega_{{\rm b}0}$, are more important. As this is
the case $N=48000$, our analysis is consistent with the faster
transport in velocity observed for the smaller values of $N$ and
thus with the more rapid thermalization in this case.

Considering the later time interval $T_2$ (Fig.~\ref{fig008}), it
is striking to note that the differences between both spectra are
strongly reduced, so that one can estimate that a test particle
feels a similar amount of stochasticity for both values of $N$.
Moreover, the relative intensity of the secondary resonances is
smaller than in the time interval $T_1$, so that the separatrix
pulsations will be relatively smaller. As a result, the chaotic
growth of the wave also slows down. By contrast, the Vlasov
simulations, with their induction of coarse graining, even prevent
the evolution of the system towards this second stage. Actually,
these codes are unable to capture the spatial details of the phase
space filamentation which are under the scale of the mesh size, so
that, after a certain stage, they artificially make the system
almost integrable.

\subsection{$(x,p)$-space structures, smoothing and
numerical entropy production in the Vlasov model}

Finally, our numerical observations along with equation
(\ref{relatrap1}) enable one to describe properly the finite-$N$
effects on the thermalization process. Due to the initial
asymmetry in velocity space (the initial distribution function has
a positive slope around the wave phase velocity), heterogeneities
will exist in the wave resonance during the first nonlinear
oscillations. If $N$ is increased, the (weak) chaos available to
thermalize the system disappears, and the system is driven by the
filamentary structure which develops inside the wave potential
well. The bottom of the well is an elliptic point for the model,
around which the correlations between trajectories may decay
algebraically \cite{Meiss}, so that the filaments get thinner and
more entangled as time goes on.

The filamentation described here mixes particles in the
neighborhood of the wave resonance. Particles with velocities far
away from the resonance are weakly sensitive to the resonance,
which they essentially average off. Therefore, as $N$ becomes very
large, the system long-time evolution is dependent on initial
conditions in the neighborhood of the resonance, and
thermodynamical conclusions (relying on ergodicity in the energy
surface and basic conservation laws) no longer apply.

Now, to what extent do actual Vlasov simulations reproduce either
the kinetic equation evolution or the finite-$N$ evolution~?
Crudely speaking, kinetic simulations start to induce a bias with
respect to the Vlasov equation, because of phase space averaging,
when the distribution function exhibits structures on scales below
their grid resolution (and this is bound to happen). One might
expect that finer grids would enable one to describe more
precisely the long-time evolution of the system. However, refined
grids also reduce the coarseness of the particle distribution in
$(x,p)$ space, and inhibit the noisy separatrix crossings. This in
turn inhibits the wave second growth, which results from the
coarseness of the particle distribution for an actual finite value
of $N$.

\begin{figure}[tbp]
  \centerline{
  \psfig{figure=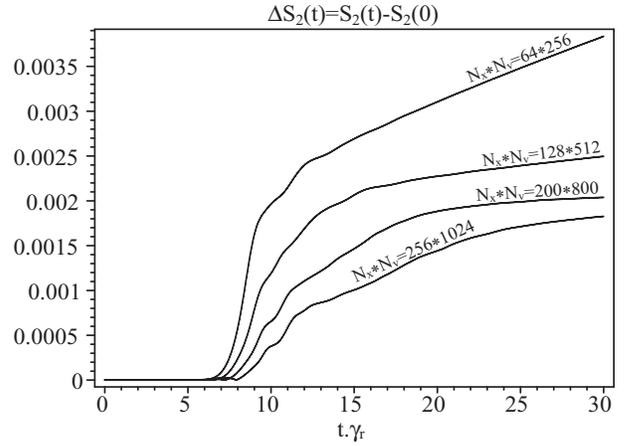,width=8cm,height=6cm}
  }
  \vskip5mm
\caption{ Evolution of the 2-entropy
  $S_2 = \int (1-f)f dx dp$
  as a function of time $\gamma_{\rm r} t$
  in the Vlasov simulations for several
  $N_x \times N_v$ grids.
  The departure from zero indicates that
  the damping of trapping oscillations is spurious.}
\label{fig009}
\end{figure}

Our observations indicate that kinetic models are too idealized,
in comparison with finite $N$, and do not contain all the
intricate behavior displayed by a discrete particles system. This
can also be tested within kinetic theory. In particular, whereas
the kinetic equation analytically preserves integral functionals
of $f$ like the 2-entropy $S_2(t) = \int (1-f)f dxdp$, numerical
schemes increase $S_2$ significantly, as shown in
Fig.~\ref{fig009}, when constant-$f$ contours form filaments in
$(x,p)$-space on scales below the grid mesh. This filamentation
(due to anisochronism of nonlinear trapped motion, or shear in
$(x,p)$-space) is smoothed by numerical partial differential
equation integrators, while $N$-body dynamics follows the
particles more realistically, sustaining the trapping
oscillations.

This inability of the semi-Lagrangian scheme in reproducing the
filamentation over small scales is shared by other Vlasov schemes.
In particular, particle-in-cell (PIC) schemes explicitly replace
the actual $N$ particles of the initial dynamics by effective
particles, which are redefined smoothly at each time step. PIC
schemes have a further disadvantage in comparison with
semi-Lagrangian schemes. They put the numerical effort on cells
according to their populations~; by contrast, semi-Lagrangian
schemes ensure a similar accuracy for the poorly populated
$(x,p)$-space domains as for the highly populated ones
\cite{Bertrand}, so that they describe frontiers in $f$ more
sharply.

This discussion shows that the `irreversible' growth of the wave
is not related to the entropy production of this sub-grid-scale
filamentation but to the chaotic trapping-detrapping process (of
which some small-scale structures are by-products). Although the
smoothing may appear as a minor nuisance in the chaotic $(x,p)$
regions, it does actually force a distinctly different evolution
in the long term, and refining the mesh does not prevent this.

\section{Comments and conclusion}

In summary, dealing with the basic propagation of a single
electrostatic wave in a warm plasma, we presented finite-$N$
effects which do not merely result from numerical errors and are
missed in a kinetic simulation approach. Their understanding
depends crucially on describing the particle dynamics in $(x,p)$
space. The sensitive dependence of the microscopic chaotic
evolution to the fine structure of the initial particle
distribution in $(x,p)$ space \cite{Firpo99} implies that the
limits $t \to \infty$ and $N \to \infty$ do not commute. The
driving process in the system evolution is separatrix crossing,
which requires a geometric approach to the system dynamics.
Further work in this direction will also shed new light on the
foundations of frequently used approximations, such as replacing
original dynamics (\ref{phi})-(\ref{force}) by coupled stochastic
equations, in which particles undergo noisy transport.

%
%
\acknowledgments

The authors are grateful to D.F.~Escande for fruitful discussions
and a critical reading of the manuscript. MCF thanks
D.~del-Castillo-Negrete and B.~Carreras for stimulating
discussions. MCF, MP and DG were supported by the French
Minist{\`{e}}re de la Recherche et de la Technologie. MCF thanks
the French Minist\`{e}re des Affaires Etrang\`{e}res for support
at the Universit\`{a} degli Studi di Firenze through a Lavoisier
fellowship. This work is part of the CNRS GdR {\it Syst{\`{e}}mes
de particules charg{\'{e}}es} SParCh.

\subsection{Appendix : Periodic orbits of the pendulum and
resonance overlap}

The pendulum equations reduce to the normal form
(\ref{mouvpartest}) with fixed parameters $\theta = 0$ and
$\omega_{{\rm b}0}$. The orbits of the pendulum in $(x,p)$ space
(with $x$ modulo $2 \pi$) are : fixed points $(0,0)$ and
$(\pi,0)$, the two branches of the separatrix and the three types
of periodic orbits. They are parametrized by the energy $E = p^2 /
2 + \omega_{{\rm b}0}^2 (1 - \cos x) \geq 0$ and (if untrapped) by
the sign of $p$. At the unstable fixed point and on the
separatrix, $E = 2 \omega_{{\rm b}0}^2$.

The separatrix branches are limits of periodic orbits with periods
going to $\infty$. Their equation $p_\pm = \pm 2 \omega_{{\rm b}0}
\cos (x/2)$ implies that the velocity ranges in $[- 2 \omega_{{\rm
b}0}, 2 \omega_{{\rm b}0}]$.

Circulating orbits have periods decreasing for increasing energy
$E = 2 \omega_{{\rm b}0}^2 k^{-2}$, where $0 < k < 1$. Their
period is $T(k) = 2 k \omega_{{\rm b}0}^{-1} {\rm K}(k^2)$, with
the complete elliptic integral K, and $k \to 1$ on approaching the
separatrix. Period $T = 2 \pi \omega_{{\rm b}0}^{-1}$ occurs for
$k \approx 0.99$, i.e. extremely close to the separatrix, with $E
\approx 2.04~\omega_{{\rm b}0}^2$. The next strong resonance, with
$T = \pi \omega_{{\rm b}0}^{-1}$, occurs for $k \approx 0.8$, i.e.
for particles with $E \approx 3.17~\omega_{{\rm b}0}^2$ and
$1.5~\omega_{{\rm b}0} \leq |p| \leq 2.5~\omega_{{\rm b}0}$.

Trapped orbits have periods $T(k) = 4 \omega_{{\rm b}0}^{-1} {\rm
K}(k^2)$, with $0 < k < 1$. The period is larger than $2
\pi/\omega_{{\rm b}0}$ and increases with the energy $E = 2
\omega_{{\rm b}0}^2 k^2$. The velocity is in the range
$[-\sqrt{2E}, \sqrt{2E}]$.

To apply the resonance overlap picture, one considers only the
relative velocity of the two waves whose resonances overlap. As
the range associated with the main cat's eye is $[- 2 \omega_{{\rm
b}0}, 2 \omega_{{\rm b}0}]$, the classical resonance overlap
picture makes sense only for waves with relative velocity larger
than $2 \omega_{{\rm b}0}$ with respect to the principal one.



\clearpage

\clearpage

\end{document}